\documentclass[conference]{IEEEtran}

\IEEEoverridecommandlockouts

\usepackage[caption=false,font=normalsize,labelfont=sf,textfont=sf]{subfig}
\usepackage{textcomp}
\usepackage{stfloats}
\usepackage{verbatim}
\usepackage{cite}
\usepackage{amsmath,amssymb,amsfonts}
\usepackage{nicefrac}
\usepackage{algorithmic}
\usepackage{graphicx}
\usepackage{float}
\usepackage{textcomp}
\usepackage{xcolor}
\usepackage{bbm}
\usepackage{array}
\usepackage{rotating}
\usepackage{gensymb}
\usepackage{comment}
\usepackage[acronym]{glossaries}
\usepackage{multirow}
\usepackage{lscape}
\usepackage{makecell}
\usepackage[inline]{enumitem}
\usepackage{siunitx}

\usepackage{tikz}
\usepackage{tikz-3dplot}
\usetikzlibrary{math}
\usetikzlibrary{backgrounds} 
\usetikzlibrary{positioning, shapes.geometric}
\usetikzlibrary{arrows.meta}
\usetikzlibrary{patterns}
\usetikzlibrary{decorations.pathmorphing}
\usepackage{pgfplots}
\pgfplotsset{compat=1.16}
\usepgfplotslibrary{patchplots}
\usepgfplotslibrary{fillbetween}
\usepgfplotslibrary{groupplots}
\definecolor{myblue}{rgb}{0.00000,0.44700,0.74100}%
\definecolor{mygreen}{rgb}{0.46600,0.67400,0.18800}%
\definecolor{amber}{rgb}{1.0, 0.75, 0.0}
\definecolor{lightgray}{rgb}{0.83, 0.83, 0.83}
\definecolor{darkgray}{rgb}{0.4, 0.4, 0.4}
\definecolor{myred}{rgb}{0.77, 0.12, 0.23}

\usepackage{hyperref}
\usepackage{cleveref}

\usepackage{adjustbox}
\usepackage{tikz}
\usetikzlibrary{shapes,arrows}
\usetikzlibrary{positioning}
\usetikzlibrary{decorations.text}

\newcommand{\ones}[1]{\mathbbm{1}\left(#1\right)}

\renewcommand{\exp}[1]{e^{#1}}

\renewcommand{\d}{\text{d}}

\newcounter{CorrCounter}
\newcounter{LemmaCounter}
\newcounter{PropCounter}

\newcommand{\pa}[1]{\left(#1\right)}
\newcommand{\abs}[1]{\left|#1\right|}

\def\BibTeX{{\rm B\kern-.05em{\sc i\kern-.025em b}\kern-.08em
    T\kern-.1667em\lower.7ex\hbox{E}\kern-.125emX}}

\begin{document}

\title{Near-Field EM-Based Multistatic Radar Range Estimation}

\author{\IEEEauthorblockN{François De Saint Moulin and Guillaume Thiran, Christophe Craeye, Luc Vandendorpe, Claude Oestges\thanks{François De Saint Moulin and Guillaume Thiran are Research Fellows of the Fonds de la Recherche Scientifique - FNRS.}}\\
\IEEEauthorblockA{ICTEAM, UCLouvain - Louvain-la-Neuve, Belgium\\}
\{francois.desaintmoulin,guillaume.thiran,christophe.craeye,luc.vandendorpe,claude.oestges\}@uclouvain.be}

\maketitle

\thispagestyle{plain}%to add vs delete page numbers: plain vs empty
\pagestyle{plain}

%\pagenumbering{gobble}

%%%%%%%%%%%%%%%%%%%%%%

\begin{abstract}
Radar targets are traditionally modelled as point target reflectors, even in the near-field region. Yet, for radar systems operating at high carrier frequencies and small distances, traditional radar propagation models do not accurately model the scatterer responses. In this paper, a novel electromagnetic-based model is thus developed for the multistatic radar detection of a rectangular plate reflector in the near-field region. This model is applied to an automotive scenario, in which a linear antenna array is spread out at the front of a vehicle, and performs a radar measurement of the distance to the back of the vehicle ahead. Based on the developed received signal model, the maximum likelihood estimator of the range is designed. By exploiting the near-field target model, this estimator is shown to provide a significant gain with respect to traditional range estimators. The impact of the system and scenario parameters, i.e. the carrier frequency, bandwidth and distance to the target, is furthermore evaluated. This analysis shows that the radar resolution in the near-field regime is improved at high carrier frequencies, while saturating to the traditional bandwidth-dependent resolution in the far-field region. 
\end{abstract}


%%%%%%%%%%%%%%%%%%%%%%

\section{Introduction}
\label{sec:introduction}
Emerging radar systems are envisioned to utilise  high carrier frequencies, as for instance the 77-81 GHz band for automotive applications \cite{ETSI}. The spectrum availability in these high frequency regimes enables to provision large bandwidths, and thus to improve the sensing performance. Such improvement is also provided by the increasing dimensions of the antenna arrays, and in particular by extremely large antenna arrays \cite{2307.02684v2}. Owing to the high carrier frequencies and large antenna arrays, the Near-Field (NF) propagation region is enlarged. It is indeed delimited by the Fraunhofer distance, which grows with the square of the array size, and with the carrier frequency. In radar applications, the NF propagation effects also depend on the target dimensions. While traditional Far-Field (FF) analyses consider point targets, targets with significant dimensions should be modelled in accordance with NF the propagation conditions. Applications such as automotive and indoor sensing thus call for accurate NF radar propagation models, and signal processing algorithms leveraging on them.

\subsection{Related works}
The NF propagation region raises both challenges and opportunities, which have been studied both for communication and radar systems   \cite{09903389,2307.02684v2,2305.17751}. On the one hand, considering the NF propagation as a challenge, \cite{09770180} proposes a strategy for compensating the range-dependent phase deviation in NF coherent radar networks in order to increase the angle-estimation capabilities of such system. On the other hand, considering the NF propagation as an opportunity, the performance achieved by a large-scale radar array is studied \cite{09707730,09743499}, when detecting a target in NF with an Orthogonal Frequency Division Multiplexing (OFDM) waveform. By leveraging on the NF propagation, the authors show numerically and experimentally that the estimation of the position can be improved with respect to traditional FF propagation models. To evaluate the performance improvement brought by NF algorithms, the Cramér-Rao Bound (CRB) has been evaluated in different scenarios. In \cite{wang2024cramer, sakhnini2021cramer}, radar passive targets are localised, the accuracy being linked to waveform parameters.  However, the above works all consider point targets, the NF effect only coming from the antenna array dimensions. In order to consider targets with significant dimensions, it is nonetheless necessary to develop novel propagation models based on ElectroMagnetic (EM) theory, to accurately model the NF effects brought by the targets themselves. This is done in \cite{10147356}, where a point antenna position is estimated from the signal received over a large surface, relying on a specific EM propagation model. Nevertheless, this work does not consider the detection of a passive radar target.

To our best knowledge, radar systems have never consider the detection of passive reflective targets, in which both the antenna arrays and the targets have significant dimension. This thus calls for i)~the development of an EM signal model compatible with such setting, and ii)~the design of the associated localisation estimators, as well as the study of their performance. 

\subsection{Contributions}
The contributions of this paper are summarised as follows:
\begin{itemize}
\item A novel EM-based signal model is developed for a rectangular plate with significant dimensions in the NF, and validated numerically. This model describes the signal received for each pair of antennas, revealing the dependence of the received signal on the specular point position.
\item Leveraging on the developed model, a maximum likelihood estimator is obtained for the distance between the antenna array and the target. Multiple levels of target knowledge are studied, in order to uncouple the estimator from the specific target geometry.
\item The designed estimator is applied to an automotive scenario, and the impact of the bandwidth, carrier frequency and target distance is analysed. For close ranges in NF, the achieved performance shows that the radar resolution does not increase only with the bandwidth, but also with the carrier frequency.
\end{itemize}

\subsection{Structure of the paper}

First, the EM-based signal model for a rectangular plate in NF is developed in \Cref{sec:em_development}, the associated maximum likelihood estimator of the target distance being obtained in \Cref{sec:ml_development}. Finally, numerical analyses and validations are performed for an automotive scenario in \Cref{sec:numerical_analysis}.

\section{Electromagnetic-based Reflection on a Square Plate}
\label{sec:em_development}
\subsection{System model}

\begin{figure}
    \centering
    \includegraphics[width=0.8\linewidth]{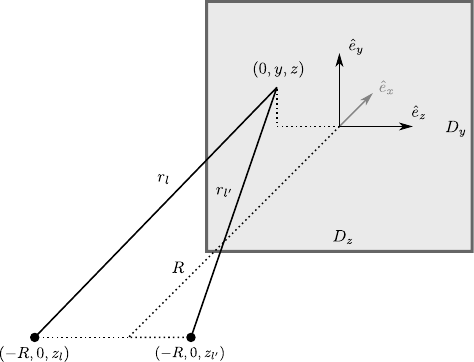}
    \caption{Reflection on a square plate from a linear array.}
    \label{fig:square_plate}
\end{figure}

Let us consider the scenario illustrated in Figure \ref{fig:square_plate}. A rectangular plate of width $D_z$ and height $D_y$ is located at a distance $R$ of $N$ antennas, distributed linearly in $\mathcal{A}=\{(x,y,z) \:|\: x=-R,y=0\}$. Focusing on one pair of antennas, the position of the transmit and receive antennas are respectively denoted by $(-R,0,z_l)$ and $(-R,0,z_{l'})$. Assuming Hertzian dipoles axed along $\hat{e}_y$, the transmitted electric field at a distance $r_l$, elevation angle $\theta_l$, azimuth angle $\psi_l$ and time $t$ is computed as \cite{orfanidis2016electromagnetic}
\begin{align}
    \vec{E}_{l}^{i}(t) &= -jk\eta L I_0 \: s_l\left(t-\frac{r_l}{c}\right) \: \frac{\exp{-jkr_l}}{4\pi r_l} \: \cos(\theta_l) \: \hat{e}_{\theta_l}, %\\
   % \vec{H}_{l}^{i} &= \frac{1}{\eta} \: \hat{e}_{r_l} \times \vec{E}_{l}^{i} = -jk L I_0 \: s_l\left(t-\frac{r_l}{c}\right) \: \frac{\exp{-jkr_l}}{4\pi r_l} \: \cos(\theta_l) \: \hat{e}_{\phi_l},
\end{align}
with $k=2\pi/\lambda$ the wave number, $\lambda$ the wavelength, $\eta$ the free-space impedance, $L$ the antenna length, $I_0$ the current in the antenna, $s_l(t)$ the signal transmitted by antenna $l$, and $\hat{e}_{r_l}$, $\hat{e}_{\theta_l}$, $\hat{e}_{\phi_l}$ the radial, elevation and azimuth unit vectors in a polar coordinate system centred on antenna $l$. The transmit signal is assumed to have a bandwidth $B \ll c/\lambda$. These equations are valid when $r_l \gg \lambda$, which is always the case at the considered carrier frequencies for automotive radar systems. 

The electrical current density generated at the surface of the plate is computed at every point of the plate from the magnetic field $\vec{H}^i_l(t) =   \hat{e}_{r_l} \times \vec{E}_{l}^{i}(t)/\eta  $ as 
\begin{align}
    \vec{J}_l(t) &= 2 \: \left(-\hat{e}_x\right) \times \vec{H}^i_l(t), \\
    &= - j 2 k L I_0  \: s_l\left(t-\frac{r_l}{c}\right) \: \frac{\exp{-jkr_l}}{4\pi r_l} \: \cos(\theta_l) \: \cos(\phi_l) \: \hat{e}_y.
    \nonumber 
\end{align}
 The noise-free received signal is finally computed from the electrical current density as \cite{orfanidis2016electromagnetic}
\begin{align}
    u_{l'l}(t) = -jk\eta \iint_{\mathcal{A}} \left(\vec{J}^\perp_{l'l}\left(t-\frac{r_{l'}}{c}\right) \cdot \vec{f}^\perp_{l} \right) \: \frac{\exp{-jkr_{l'}}}{4\pi r_{l'}} \: \d\mathcal{A}, \label{eq:backscattered_signal_first}
\end{align}
with $\vec{J}^\perp_{l'l}(t) \triangleq  \vec{J}_{l}(t) - \left(\vec{J}_{l}(t) \cdot \hat{e}_{r_{l'}}\right)\hat{e}_{r_{l'}}$ the component of the surface current density perpendicular to the direction of propagation, and $\vec{f}^\perp_{l} \triangleq  L \cos\theta_{l'} \: \hat{e}_{\theta_{l'}}$ the antenna effective length .
%\begin{align}
%    &\vec{J}^\perp_{l'l} = \vec{J}_{l} - \left(\vec{J}_{l} \cdot \hat{e}_{r_{l'}}\right)\hat{e}_{r_{l'}} = - j2 kL I_0   \\
%    &\qquad \cdot s_l\left(t-\frac{r_l+ r_{l'}}{c}\right) \: \frac{\exp{-jkr_l}}{4\pi r_l} \:\cos(\theta_l) \: \cos(\phi_l) \: \cos^2(\theta_{l'}). 
%\end{align}
Developing the scalar product, \eqref{eq:backscattered_signal_first} is rewritten as 
\begin{equation}
u_{l'l}(t) = -\frac{2 k^2\eta L^2 I_0}{(4\pi)^2} \int_{-\frac{D_z}{2}}^\frac{D_z}{2} \int_{-\frac{D_y}{2}}^\frac{D_y}{2} g_{l'l}(y,z) \: e^{j\psi_{l'l}(y,z)}\: \d y \d z, \label{eq:backscattered_signal} 
\end{equation}
with 
\begin{align}
g_{l'l}(y,z) &= s\left(t-\frac{r_l+ r_{l'}}{c}\right) \: \frac{\cos(\theta_l) \: \cos(\phi_l) \: \cos^2(\theta_{l'})}{r_l r_{l'}} , \label{eq:amplitude_integrand}\\
\psi_{l'l}(y,z) &= -k(r_l + r_{l'}). \label{eq:phase_integrand}
\end{align}

\subsection{Stationary phase approximation}
Let us consider an integral of the form
\begin{equation}
    I = \int g(x) \: e^{j\psi(x)} \: \d x,
\end{equation}
where $g$ and $\psi$ are amplitude and phase functions. The Stationary Phase Approximation (SPA) helps to solve this integral by approximating the function $\psi$ with a Taylor expansion around a stationary point $x_s$, defined such that $\d\psi/\d x |_{x = x_s} = 0$ \cite{stationary_phase}:
\begin{equation}
    \psi(x) = \psi(x_s) + \frac{1}{2} \left.\frac{\d^2 \psi}{\d x^2}\right|_{x=x_s} (x-x_s)^2 + \mathcal{O}(x^3).
\end{equation}
Assuming that the function $g$ varies slowly in the vicinity of $x_s$ for which the function $\psi$ does not oscillate quickly, denoting by $\psi''_{xs}$ the second derivative of $\psi$ around $x_s$, the integral is approximated as 
\begin{align}
    I &\approx g(x_s) \: e^{j\psi(x_s)} \int e^{j \frac{\psi''_{xs}}{2} (x-x_s)^2} \: \d x.
\end{align}
This has been illustrated in one dimension for sake of clarity, but it can be expanded to multivariate integrals.

\subsection{Application to the considered model}

In the considered scenario, the varying phase of the integrand in \eqref{eq:backscattered_signal} at $(0,y,z)$ on the plate is given by \eqref{eq:phase_integrand}, with
\begin{equation}
    r_l+r_{l'} = \sqrt{R^2 + y^2 + (z-z_l)^2} + \sqrt{R^2 + y^2 + (z-z_{l'})^2}. \label{eq:r_l}
\end{equation}
In this case, the stationary point is also the specular point,  with coordinates given by
\begin{equation}
y_{l'l} \triangleq 0,\qquad z_{l'l} \triangleq \frac{z_l + z_{l'}}{2}.
\end{equation}
To distance to the specular point is defined as $r_{l'l} \triangleq r_l(y_{l'l},z_{l'l}) = r_{l'}(y_{l'l},z_{l'l})$. The SPA then requires to approximate the phase function \eqref{eq:phase_integrand} with a Taylor expansion as 
\begin{equation}
    \psi_{l'l}(y,z) \approx -2k\:r_{l'l} - \frac{k}{r_{l'l}} y^2 - \frac{k R^2}{r_{l'l}^3}\left(z-z_{l'l}\right)^2,\label{eq:phase_SPA}
\end{equation}
and to evaluate \eqref{eq:amplitude_integrand} at the stationary point as 
\begin{equation}
    g_{l'l}(0,z_{l'l}) = s_l\left(t-\frac{2 r_{l'l}}{c}\right) \: \frac{R}{r_{l'l}^3}. \label{eq:amp_SPA}
\end{equation}
Together, \eqref{eq:phase_SPA} and \eqref{eq:amp_SPA} enable to rewrite \eqref{eq:backscattered_signal} as 
\begin{equation}
    u_{l'l}(t) \approx \beta_{l'l} \: s_l\left(t-\frac{2r_{l'l}}{c}\right),
    \label{eq:signal_model}
\end{equation}
with the complex coefficient $\beta_{l'l}$ reading as 
\begin{equation}
    \beta_{l'l} = \frac{-2 k^2\eta L^2 I_0}{(4\pi)^2} \: \frac{R}{r_{l'l}^3} \: \exp{-j2k r_{l'l}} \: S_{l'l} \: \ones{\left|z_{l'l}\right| \leq \frac{D_z}{2}}, \label{eq:alpha_inter}
\end{equation}
and with $S_{l'l}$ defined as
\begin{align}
    &S_{l'l} = \int_{-D_y/2}^{D_y/2} \exp{-j\frac{k}{r_{l'l}}y^2} \d y \cdot \int_{-D_z/2}^{D_z/2} \exp{-j\frac{k R^2}{r_{l'l}^3}(z-z_{l'l})^2} \d z.
\end{align}
Letting the Fresnel integral be $F(x) \triangleq \int_{0}^x e^{j\pi \frac{t^2}{2}} \: \d t$, the received signal is finally obtained as 
\begin{equation}
    u_{l'l}(t) = \xi \: \alpha_{l'l} \: \frac{e^{-j2kr_{l'l}}}{r_{l'l}} \: s_{l}\left(t-\frac{2r_{l'l}}{c}\right), \label{eq:backscattered_signal_final}
\end{equation}
with $\xi = - k \eta L^2 I_0 / 8 \pi$, and $\alpha_{l'l}$ defined in \eqref{eq:alpha_final}.
\begin{figure*}
%\hrule
\begin{align}
%    \nonumber S_{l'l} &= \sqrt{\frac{2\pi r_{l'l}}{k}}F^*\left(\sqrt{\frac{2k}{\pi r_{l'l}}} \frac{D_y}{2}\right) \cdot \sqrt{\frac{\pi r_{l'l}^3}{2k R^2}}\left[F^*\left(\sqrt{\frac{2kR^2}{\pi r_{l'l}^3}}\left(\frac{D_z}{2}- z_s\right)\right) - F^*\left(\sqrt{\frac{2kR^2}{\pi r_{l'l}^3}}\left(-\frac{D_z}{2}- z_s\right)\right)\right],\\
%    &= \frac{\pi r_{l'l}^2}{k R} F^*\left(\sqrt{\frac{D_y^2}{\lambda r_{l'l}}}\right) \left[F^*\left(\sqrt{\frac{(D_z-z_l-z_{l'})^2 R^2}{\lambda r_{l'l}^3}}\right) + F^*\left(\sqrt{\frac{(D_z+z_l+z_{l'})^2 R^2}{\lambda r_{l'l}^3}}\right)\right].\label{eq:surface}\\
     \alpha_{l'l} &= F^*\left(\sqrt{\frac{D_y^2}{\lambda r_{l'l}}}\right) \left[F^*\left(\sqrt{\frac{(D_z-z_l-z_{l'})^2 R^2}{\lambda r_{l'l}^3}}\right) + F^*\left(\sqrt{\frac{(D_z+z_l+z_{l'})^2 R^2}{ \lambda r_{l'l}^3}}\right)\right] \ones{\left|z_{l'l}\right| \leq \frac{D_z}{2}}. \label{eq:alpha_final}
\end{align}
\hrule
\end{figure*}
%It is finally solved in \eqref{eq:surface} using
%\begin{equation}
%    \int_{a}^b e^{jc x^2}\: \d x =\sqrt{\frac{\pi}{2c}} \left[F\left(\sqrt{\frac{2c}{\pi}} \: b\right) - F\left(\sqrt{\frac{2c}{\pi}} \: a\right)\right],
%\end{equation}
%with $F$ being the Fresnel integral:
%\begin{equation}
%    F(x) = \int_{0}^x e^{j\pi \frac{t^2}{2}} \: \d t,
%\end{equation}
%and knowing that $F(-x) = -F(x)$.  \smallskip

%\begin{figure}
%    \centering
%    \includegraphics[width=0.9\linewidth]{Images/fresnel.pdf}
%    \caption{Fresnel function $F$. The dashed line represents the value $x=\sqrt{D_y^2/\lambda R}$ with parameters of Table \ref{tab:parameters}.}
%    \label{fig:fresnel}
%\end{figure}

%To summarise, thanks to the EM theory, the received signal for each pair of antennas is expressed as 
%\begin{equation}
%    u_{l'l}(t) = \xi \: \alpha_{l'l} \: \frac{e^{-j2kr_{l'l}}}{r_{l'l}} \: s_{l}\left(t-\frac{2r_{l'l}}{c}\right), \label{eq:backscattered_signal_final}
%\end{equation}
%with $\xi = -\rho k \eta L I_0 / 16 \pi$, and $\alpha_{l'l}$ defined in \eqref{eq:alpha_final}. 
While the coefficient $\xi$, identical for all antenna pairs,  is standard in radar NF models, these latter lack the three following NF effects when targets with non-negligible dimensions are considered:
\begin{enumerate}
\item the antenna-dependent coefficient $\alpha_{l'l}$, non-zero only if the specular point is located on the target;
\item the fact that the phase shift and amplitude decay depend on the distance between the transmit or receive antenna and the specular point $r_{l'l}$, different for each antenna pair;
\item the transmit signal delay $2 r_{l'l}/c$, being also dependent on the distance between the transmit or receive antenna and the specular point.
\end{enumerate}
The distance of the target can therefore be extracted from the delay of the transmitted signal as usually done in traditional radar systems, but also from the additional antenna-dependent attenuations and phase shifts highlighted in \eqref{eq:backscattered_signal_final}. This implies that the resolution limit of $c/2B$ reached by traditional processing could be outreached with high carrier frequencies \cite{radar_fund}, as formalised in Section \ref{sec:ml_development}.

\section{Range Estimation}
\label{sec:ml_development}
Leveraging on the EM signal model developed in Section \ref{sec:em_development}, this section focuses on the development of a new estimator of the range $R$. The received signal $u_{l'}$ of antenna $l'$ is the combination of the transmitted signals from all the antennas:  $u_{l'}(t) = \sum_{l=0}^{N-1} u_{l'l}(t) + w_{l'}(t)$, where $w_{l'}$ are independent Additive White Gaussian Noise (AWGN) at each receive antenna. According to \eqref{eq:backscattered_signal_final}, the received signal reads as 
\begin{align}
    u_{l'}(t) = \xi \: \mu_{l';R}\pa{t}+ w_{l'}\pa{t}, \label{eq:signal_full}
\end{align}
with $\xi$ an unknown complex coefficient identical for all the antennas. The function $\mu_{l';R}$ is defined as 
\begin{equation}
\mu_{l';R}(t) = \sum_{l=0}^{N-1} \frac{e^{-j2kr_{l'l}}}{r_{l'l}} \: \alpha_{l'l} \: s_l \pa{t-\frac{2r_{l'l}}{c}}.
\label{eq:mu_model}
\end{equation}
As a reminder, the range $R$ of the target impacts both the signal delay $2r_{l'l}/c$, and the complex coefficients related to each pair of antennas. The maximum likelihood estimators of $R$ and $\xi$ are obtained as
\begin{align}
    \hat{R},\hat{\xi} &= \arg\min_{R,\xi} \sum_{l'=0}^{N-1} \int_t\left|u_{l'}(t) - \xi \: \mu_{l';R}\pa{t} \right|^2 \d t. \label{eq:ml_full}
\end{align}
Developing the 2-norm, $\hat{\xi}\pa{R}$ is expressed as:
\begin{equation}
    \hat{\xi}(R) = \frac{\sum_{l'=0}^{N-1} \int_t u_{l'}\pa{t} \mu_{l';R}^*\pa{t}  \mathrm{d}t}{\sum_{l'=0}^{N-1}\int_t \abs{\mu_{l';R}\pa{t}}^2 \mathrm{d}t }.
\end{equation}
Inserting the above into \eqref{eq:ml_full}, $\hat{R}$ is computed as 
\begin{equation}
    \hat{R} = \arg\max_R \: \Lambda(R).
    \label{eq:ML_estimator}
\end{equation}
with the modified likelihood function obtained as 
\begin{equation}
    \Lambda(R) \triangleq  \frac{\abs{\sum_{l'=0}^{N-1} \int_t u_{l'}\pa{t} \mu_{l';R}^*\pa{t}  \mathrm{d}t}^2}{\sum_{l'=0}^{N-1}\int_t \abs{\mu_{l';R}\pa{t}}^2 \mathrm{d}t }.
    \label{eq:ML_fct}
\end{equation}
If orthogonal resources are utilised for the different signals $s_l\pa{t}$, a noisy version of each individual signal $u_{l'l}(t)$ of \eqref{eq:backscattered_signal_final} is available at the receiver for all pairs $l,l'$. Letting $\mu_{l'l;R}\pa{t}\triangleq \frac{e^{-j2kr_{l'l}}}{r_{l'l}} \: \alpha_{l'l} \: s_l \pa{t-\frac{2r_{l'l}}{c}}$, the above becomes in that case:
\begin{equation}
    \Lambda(R) \triangleq  \frac{\abs{\sum_{l'=0}^{N-1}\sum_{l=0}^{N-1} \int_t u_{l'l}\pa{t} \mu_{l'l;R}^*\pa{t}  \mathrm{d}t}^2}{\sum_{l'=0}^{N-1}\sum_{l=0}^{N-1}\int_t \abs{\mu_{l'l;R}\pa{t}}^2 \mathrm{d}t }.
    \label{eq:ML_fct_orth}
\end{equation}

\begin{figure}
    \centering
    \includegraphics[width = 0.8\columnwidth]{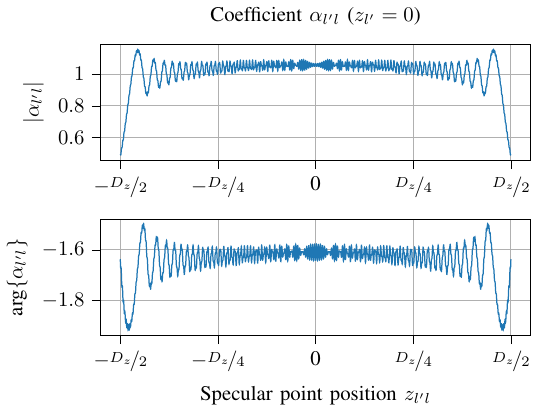}
    \caption{Coefficient $\alpha_{l'l}$ of \eqref{eq:alpha_final} as a function of the specular point position, evaluated with parameters of Table \ref{tab:parameters} and $z_{l'} = 0$.}
    \label{fig:alpha}
\end{figure}

Note that multiple models of $\mu_{l';R}$ can be considered depending on the level of knowledge about the target. In this paper, either a full or a partial knowledge of the target are considered. In the first case, the complete expression of $\alpha_{l'l}$  \eqref{eq:alpha_final} is taken into account. As this expression heavily depends on the target geometry, the almost constant behaviour of the  Fresnel function $F\pa{x}$ for large $x$'s can be leveraged upon. As depicted in \Cref{fig:alpha}, this enables to consider coefficients $\alpha_{l'l}$ independent of the specular position, and thus not varying for each pair antenna, as long as the specular point is not too close to the plate edges. This leads to $   \alpha_{l'l} \approx \alpha  \ones{\left|{z_{l'l}}\right| \leq \nicefrac{D_z}{2}}$,
with $\alpha$ a constant complex coefficient included in $\xi$ in the ML development. The simplified model is thus obtained as 
\begin{equation}
    \Tilde{\mu}_{l';R}(t) \triangleq \sum_{l=0}^{N-1} \ones{\left|z_{l'l}\right| \leq \frac{D_z}{2}} \frac{e^{-j2kr_{l'l}}}{r_{l'l}} \: s_l \pa{t-\frac{2r_{l'l}}{c}}.
\label{eq:simpl_model}
\end{equation}
This model is close to the usual received signal model used in the literature, except that the delay, attenuation and phase shifts are computed based on the position of the specular point, and not of the centre of the target.

%\begin{figure}
%    \centering
%    \includegraphics[width=0.9\linewidth]{Images/coef_alpha.pdf}
%    \caption{Coefficient $\alpha_{l'l}$ of \eqref{eq:alpha_final} against the position of the specular point, evaluated with parameters of Table \ref{tab:parameters} and $z_{l'} = 0$.}
%    \label{fig:alpha}
%\end{figure}

\section{Numerical Analysis}
\label{sec:numerical_analysis}
\begin{table}
    \centering
    \caption{Scenario's parameters}
    \label{tab:parameters}
    \renewcommand{\arraystretch}{1.1}
    \begin{tabular}{|l l|}
    \hline
        Number of antennas &$N = 13$  \\\hline
        Antenna spacing &$\Delta = 0.125$ m \\ \hline
        Antenna length/current & $L^2  I_0 = 1$ m$^2$A \\\hline
        Signal bandwidth & $B = 100$ MHz \\\hline
        Carrier frequency &$f_c = 77$ GHz \\\hline
        \multirow{2}{*}{Target dimensions} & $D_y = 0.8$ m \\
        & $D_z = 1.75$ m \\\hline
        Target Distance & $R = 4$ m \\\hline
    \end{tabular}
    \renewcommand{\arraystretch}{1.}
    \vspace*{-0.6cm}
\end{table}

In this section, a numerical validation and analysis of the EM model and ML range estimator is provided. This numerical scenario mimics an automotive setting, in which one car estimates the distance to the car driving ahead. To that aim, $N$ antennas are spread linearly at the front of the vehicle with a spacing of $\Delta$, the $z$-position of antenna $l$ being given by $z_l = \pa{-\frac{N-1}{2} + l} \Delta$. The radar signals $s_l(t)$ are sent on orthogonal time resources, with the waveforms $s(t)\triangleq \frac{\sin\pa{\pi B t}}{\pi B t} $, $B$ being the signal bandwidth. Unless stated otherwise, the scenario parameters are those of Table~\ref{tab:parameters}. 

%This numerical scenario mimics an automotive setting, in which one car estimates the distance to the car driving ahead. To that aim, it has $N=7$ antennas spread at the front with a spacing of $\Delta = \SI{0.25}{m}$, the $z$-position of antenna $l$ being given by $z_l = \pa{-\frac{N-1}{2} + l} \Delta$. The target dimension is set to $D_z=\SI{1.75}{m}$ and $D_y=\SI{0.8}{m}$. The radar signals $s_l(t)$ are all equal to the same signal $s(t)\triangleq \frac{\sin\pa{\pi B t}}{\pi B t t} $, with $B$ the signal bandwidth. Unless stated otherwise, the radar system operates at $f_c = \SI{77}{GHz}$ with a bandwidth $B = \SI{100}{MHz}$, with a target at a distance $R=\SI{4}{m}$. Noise-free signals are always considered in this numerical analysis, since this enables to better grasp the near-field benefits. Further work will focus on the noise impact, which is expected to be similar than its effect in the far-field region. 

\subsection{Validation of the electromagnetic model}
\label{sub:EM_validation}

%\begin{figure}
    %\centering
    %\includegraphics[width=0.47\linewidth]{Images/amplitude_error.pdf}
    %\vspace*{0.5cm}
    %\hspace*{-0.15cm}
    %\includegraphics[width=0.48\linewidth]{Images/phase_error.pdf}
    %\includegraphics[width=\linewidth]{Images/amplitude_phase_error.pdf}
    %\caption{Amplitude and phase errors committed by the SPA.}
    %\label{}
%\end{figure}

\begin{figure}
    \centering
    \scalebox{0.8}{% This file was created with tikzplotlib v0.10.1.
\begin{tikzpicture}

\definecolor{darkgray176}{RGB}{176,176,176}

\begin{axis}[
at = {(0.58\columnwidth,0)},
width = 0.6\columnwidth,
axis equal image,
colorbar,
colorbar style={ylabel={}},
colormap/viridis,
point meta max=2.84305789208966,
point meta min=-2.89065860716545,
tick align=outside,
tick pos=left,
title style={align=center},
title={ 
 Phase approximation error  (°) \\ \phantom{$\left[z_{ll'}\right]_{\mathrm{dB}} - \left[z_{\mathrm{SPA},ll'}\right]_{\mathrm{dB}}$}},
x grid style={darkgray176},
xlabel={Antenna element $l$},
xmin=-0.5, xmax=12.5,
xtick style={color=black},
y grid style={darkgray176},
ymin=-0.5, ymax=12.5,
ytick style={color=black},
colorbar style={
    at={(1.1,0)},
    anchor=south west,
    width=0.02\columnwidth}
]
\addplot graphics [includegraphics cmd=\pgfimage,xmin=-0.5, xmax=12.5, ymin=-0.5, ymax=12.5] {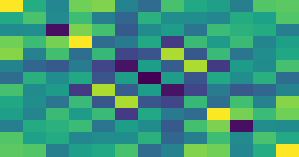};
\end{axis}

\begin{axis}[
at = {(0\columnwidth,0)},
width = 0.6\columnwidth,
axis equal image,
colorbar,
colorbar style={ylabel={}},
colormap/viridis,
point meta max=0.219171624853452,
point meta min=-0.166586063520917,
tick align=outside,
tick pos=left,
title style={align=center},
title={Amplitude approximation error \\ $\left[u_{l'l}\right]_{\mathrm{dB}} - \left[u_{\mathrm{SPA},l'l}\right]_{\mathrm{dB}}$ },
x grid style={darkgray176},
xlabel={Antenna element $l$},
xmin=-0.5, xmax=12.5,
xtick style={color=black},
y grid style={darkgray176},
ylabel={Antenna element $l'$},
ymin=-0.5, ymax=12.5,
ytick style={color=black},
colorbar style={
    at={(1.1,0)},
    anchor=south west,
    width=0.02\columnwidth}
]
\addplot graphics [includegraphics
cmd=\pgfimage,xmin=-0.5, xmax=12.5, ymin=-0.5, ymax=12.5] {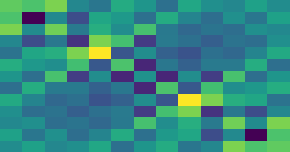};
\end{axis}

\end{tikzpicture}
    }
    \caption{Amplitude and phase errors committed by the SPA.}
    \label{fig:spa_errors}
\end{figure}

Figure \ref{fig:spa_errors} illustrates the amplitude and phase errors committed by the SPA for parameters depicted in Table \ref{tab:parameters} for a constant transmitted signal $s_l(t) = 1$. It shows differences less than 0.3dB in amplitude and 3\degree\:in phase. These low errors make the developed analytical model $\mu_{l';R}\pa{t}$ in \eqref{eq:mu_model} a precise model, based on which ML range estimation can be performed, as validated below.

\subsection{ML Ambiguity functions}
\begin{figure}[t]
    \centering
    \includegraphics[width=0.8\columnwidth]{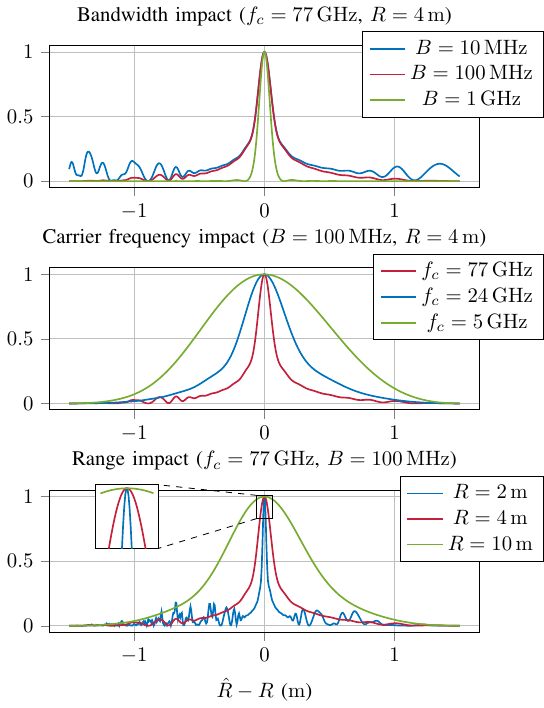}
    \caption{Impact of the parameters on the ML functions }
    \label{fig:ML_impact}
    \vspace{-0.3cm}
\end{figure}
Based on the developed EM model, the ML range estimator \eqref{eq:ML_estimator} can be analysed through its ambiguity function, i.e., the value of \eqref{eq:ML_fct} evaluated at $\hat{R}$, while the received signals are noise-free signals at a range $R$.  The impacts of the bandwidth, carrier frequency, and range on this ambiguity function are analysed in \Cref{fig:ML_impact}.  In the upper panel, one can observe that increasing the bandwidth has a twofold effect: it reduces the side lobe levels, and for high bandwidths, it reduces the main lobe width. It can be checked that for $B=\SI{1}{GHz}$, the \SI{3}{dB} pulse width is equal to the traditional $c/2B = \SI{0.15}{m}$ width. On the middle panel, the impact of the carrier frequency is observed. Increasing the carrier frequency enables to reduce the pulse width, and thus to improves the precision and resolution. For the low carrier frequency ($f_c=\SI{5}{GHz}$), the pulse width is again equal to $c/2B = \SI{1.5}{m}$. This analysis shows that the developed estimation method gives in the worst case the same performance as classical range estimation methods, while improving the estimation when the NF phase information can be utilised. The cases in which the NF effect appear depend on the range, as the bottom panel of \Cref{fig:ML_impact} shows. The closest the target is to the antenna array, the narrower the ambiguity function main lobe is.

In the above figures, the exact received signals have been obtained numerically, while the ML estimator have leveraged upon the partial-information model $\Tilde{\mu}_{l';R}\pa{t}$. This model $\Tilde{\mu}_{l';R}\pa{t}$ is an approximation of $\mu_{l';R}\pa{t}$, which is itself an analytical approximation of the exact EM propagation. Despite this, the ambiguity functions indicate the true range for all sets of parameters, the approximations being observed to have a negligible impact. Analysing the difference between the ambiguity function obtained with $\mu_{l';R}\pa{t}$ and $\Tilde{\mu}_{l';R}\pa{t}$, they have been observed to remain below $\SI{1.5}{\percent}$ for the set of parameters of \Cref{fig:ML_impact}. These low errors motivate the utilization of the simplified model \eqref{eq:simpl_model} instead of \eqref{eq:mu_model}, as the latter heavily depends on the target geometry.

\subsection{Cramér-Rao bound}

\begin{figure}[t]
    \centering
    \includegraphics[width=0.8\columnwidth]{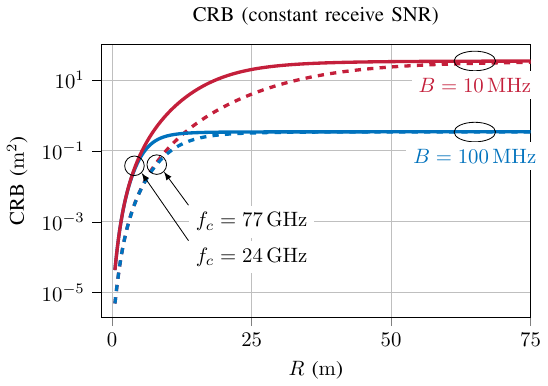}
    \caption{CRB as a function of $R$, for a constant receive SNR} 
    \label{fig:second_der}
    \vspace{-0.3cm}
\end{figure}

In order to complement the above analysis, the Cramér-Rao Bound (CRB) is numerically evaluated as the inverse of the ambiguity function second derivative. \Cref{fig:second_der} shows the CRB value\footnote{As generating the exact received signals numerically is computationally demanding, and since \Cref{sub:EM_validation} has shown that the analytical approximation has negligible differences, the coefficients $\alpha_{l'l}$ have been obtained with \eqref{eq:backscattered_signal_final} in this subsection.} as a function of the range $R$, for different carrier frequencies and bandwidths. As it can be observed, the estimation method always leverages on the most informative value among the waveform effect (related to the bandwidth) and the phase shift effect (related to the carrier frequencies). While for small ranges, the NF phase information dominates, the CRB saturates to a value depending on the bandwidth in the FF region.

\section{Conclusion}
In this paper, a novel EM-based signal model for rectangular targets in NF has been developed. Leveraging on the amplitude and phase variations of the model, a ML estimator has been designed, accommodating different knowledge levels of the target. Finally, this has been applied to an automotive scenario, and the impact of the signal carrier frequency, bandwidth and distance to the target has been analysed. It has been shown that building on the developed NF signal model, the radar performance  increases both with the signal bandwidth and with the signal carrier frequency in the NF region. In future works, other geometries of the antenna array and target will be considered. Analytical expressions of the ambiguity functions and CRBs are furthermore under study. 

\bibliographystyle{ieeetr}
\bibliography{biblio}

\appendices

\end{document}